\documentclass[reprint, amsmath,amssymb,aip]{revtex4-1}
\usepackage{graphicx}
\usepackage{natbib}
\usepackage{amsmath}
\usepackage{subfigure}
\usepackage{color}
\usepackage{dcolumn}
\usepackage{bm}
\usepackage{hyperref}

\begin{document}

\title{ Amplification and Nonlinear Mechanisms in Plane Couette Flow}

\author{Dennice F. Gayme}
\author{Beverley J. McKeon}
\affiliation{Division of Engineering and Applied Science, California
Institute of Technology, Pasadena, California, 91125}
\author{Bassam Bamieh}
\affiliation{Department of Mechanical Engineering, University of
California at Santa Barbara, Santa Barbara, California, 93106}
\author{Antonis Papachristodoulou}
\affiliation{Department of Engineering Science, University of Oxford,
Parks Road, Oxford OX1 3PJ, U.K.}
\author{John C. Doyle}
\affiliation{Division of Engineering and Applied Science, California
Institute of Technology, Pasadena, California, 91125}

\date{\today}

\begin{abstract}
We study the input-output response of a streamwise constant projection of the Navier-Stokes equations for plane Couette flow, the so-called 2D/3C model.  Study of a streamwise constant model is motivated by numerical and experimental observations that suggest the prevalence and importance of streamwise and quasi-streamwise elongated structures. Periodic spanwise/wall-normal ($z$--$y$) plane stream functions are used as input to develop a forced 2D/3C streamwise velocity field that is qualitatively similar to a fully turbulent spatial field of DNS data.  The input-output response associated with the 2D/3C nonlinear coupling is used to estimate the energy optimal spanwise wavelength over a range of Reynolds numbers.  The results of the input-output analysis agree with previous studies of the linearized Navier-Stokes equations.  The optimal energy corresponds to minimal nonlinear coupling.  On the other hand, the nature of the forced 2D/3C streamwise velocity field provides evidence that the nonlinear coupling in the 2D/3C model is responsible for creating the well known characteristic ``S'' shaped turbulent velocity profile. This indicates that there is an important tradeoff between energy amplification, which is primarily linear and the seemingly nonlinear momentum transfer mechanism that produces a turbulent-like mean profile.

\end{abstract}
\maketitle

\section{Introduction}

We study the input-output response of the Navier-Stokes (NS) equations for plane Couette flow.  Although this type of analysis can be performed for a variety of input/output combinations for the full nonlinear equations, the mathematical complexity of such an endeavor makes it difficult to both obtain and interpret the results.  Instead we perform the analysis in the simplified setting of a nonlinear streamwise constant projection of the NS equations.

The choice of a streamwise constant model is motivated by studies of the linearized Navier-Stokes (LNS) equations, which show that streamwise constant features are the dominant mode shapes that develop under various perturbations about both the laminar \cite{FI93b,BD01,JB01,JB05} and turbulent mean velocity \cite{dAJ06,McS10} profiles.  In addition, streaks of streamwise velocity naturally arise from the set of initial conditions that produce the largest energy growth \citep{BF92,FI93a}, namely streamwise vortices.  Even in linearly unstable flows, studies have shown that the amplitude of streamwise constant structures can exceed that of the linearly unstable modes \cite{JB04,G91}. Bamieh and Dahleh \cite{BD01} explicitly showed that streamwise constant perturbations produce energy growth on the order of $R^3$ whereas disturbances with streamwise variations produce
growth on the order of $R^{\frac{3}{2}}$.

The prevalence of large-scale streamwise constant structures is also supported by direct numerical simulation (DNS) and experiments.   In Couette flow, DNS has long produced turbulent flows with very large scale streamwise and quasi-streamwise structures in the core \cite{LK91,BTAlAn95}.  Experimental high Reynolds number studies have similarly identified large-scale streamwise coherence in other flow configurations~\cite{KA99,MMcJS04,GHA06,HM07_1}. At high Reynolds numbers (e.g., $Re_{\tau}>7300$), there is experimental evidence from pipes and turbulent boundary layers suggesting that these structures contain more energy than those nearer to the wall \cite{MMcJS04,HM07_1,HM07_2}.  The near-wall features, which also exhibit streamwise and quasi-streamwise alignment, are known to play a key role in energy production through the well studied ``near-wall autonomous cycle'' \cite{W90,HKW95,W97,JP99}.

The dominance of streamwise constant features was previously used to motivate the study of a streamwise constant model for plane Couette flow \cite{GMcPBD09}.  A stochastically forced version of this streamwise constant projection of NS was shown to reproduce important features of fully developed turbulence, including the shape of the turbulent velocity profile.    Further, using Taylor's hypothesis, the same model also generated large scale streaky structures, that closely resemble large-scale features in the core \cite{GMcPD09}.  Given that maximum amplification of the LNS also occurs for the $k_x=0$ modes (i.e., in a streamwise constant sense), the analysis of a streamwise constant model can be viewed as a study of the full system along the direction of maximum amplification.  This approach allows us to understand the interaction between the well studied linear amplification mechanisms and additional effects due to the nonlinear coupling in the streamwise constant model.

\begin{figure*}[!t]
\begin{center}
\subfigure[]{\label{fig:Psi-a}
\includegraphics[height=2.5 in, width=3.5 in,clip]{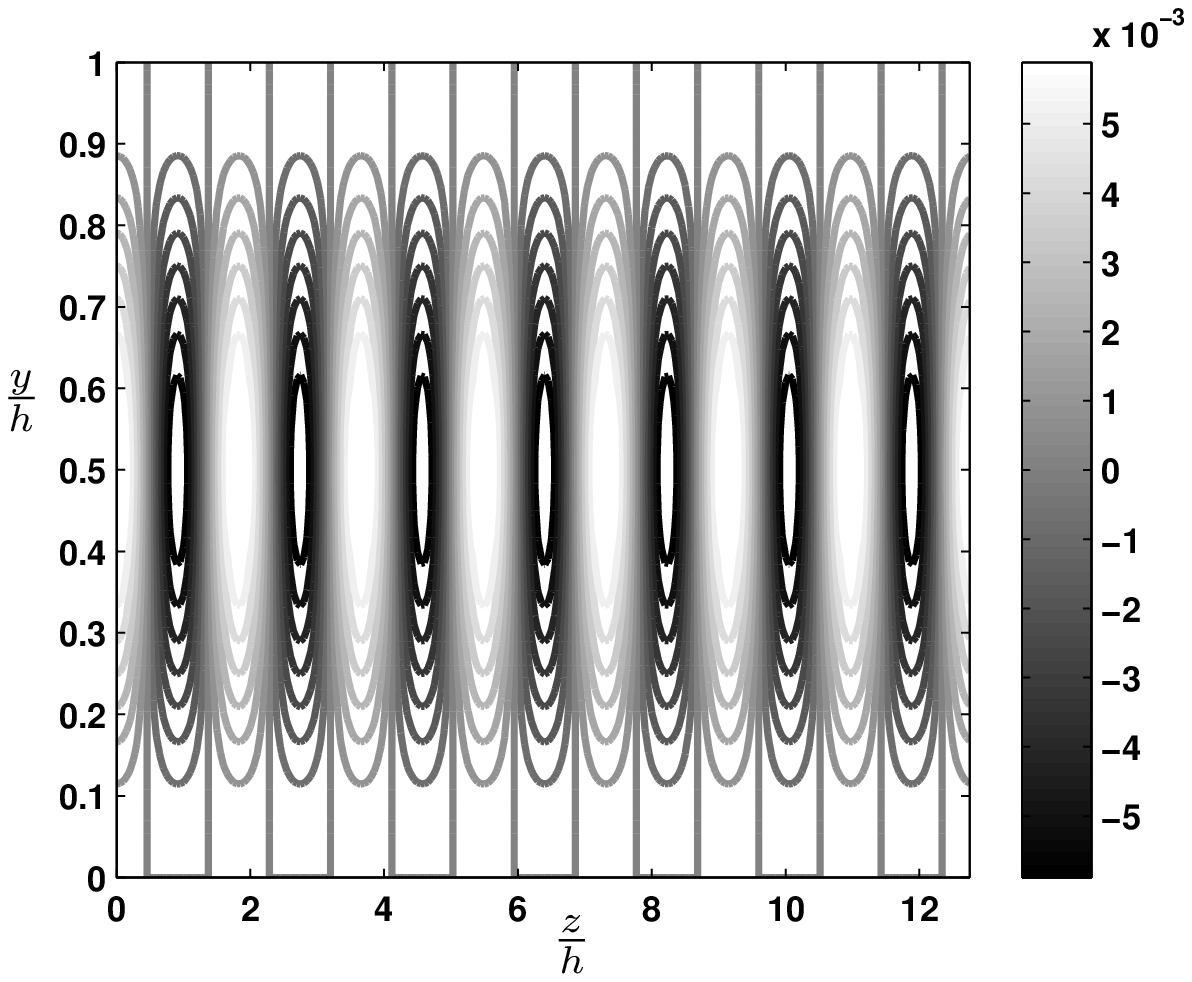}}
\subfigure[]{\label{fig:Psi_from_w0}
\includegraphics[height =2.5 in, width=3.25 in,clip]{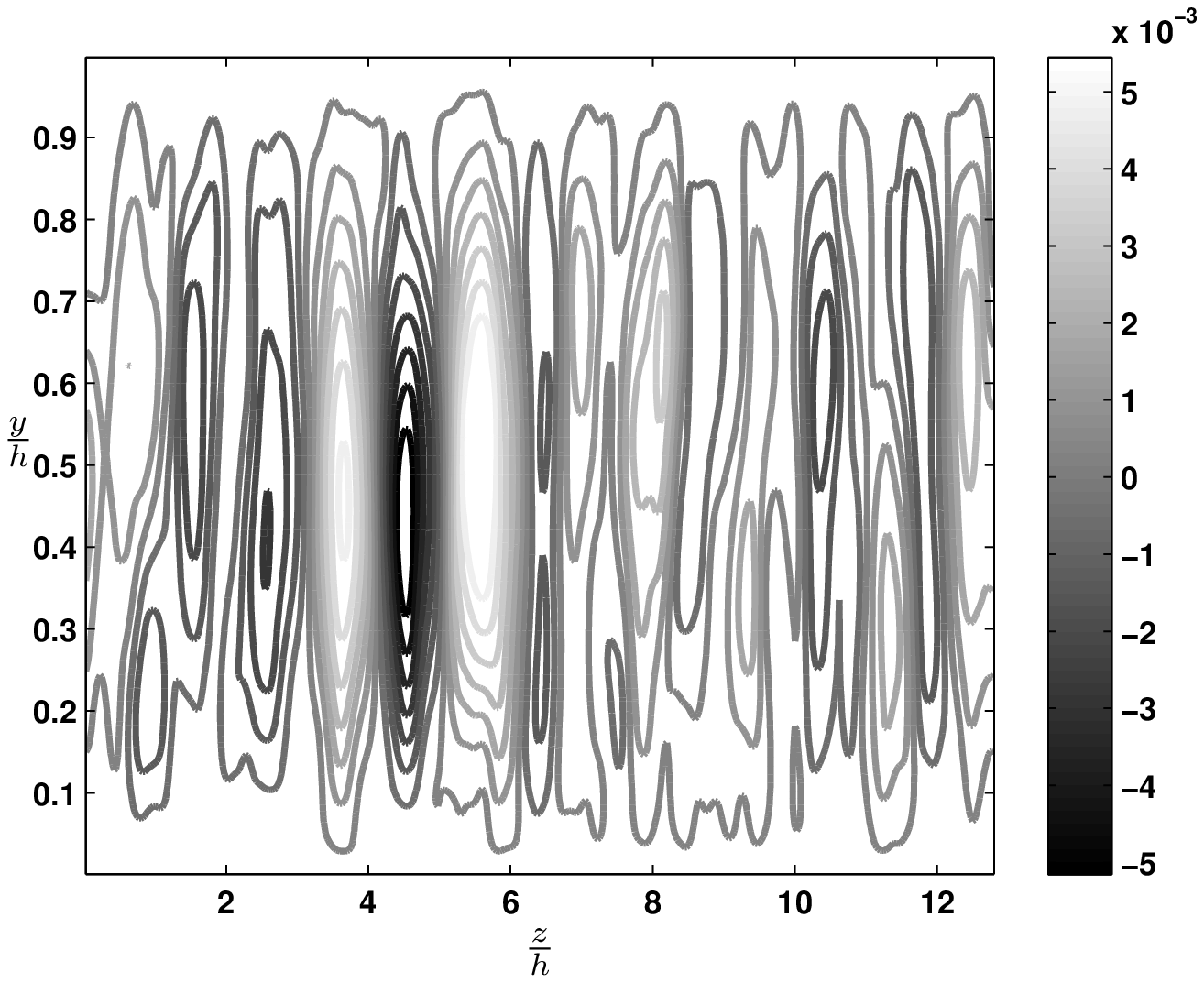}}
\caption{(a) Contour plot of the first $y$ harmonic ($q=1$)
for the stream function model.  This model represents the
streamwise constant streaks and vortices commonly observed in DNS
and experiments.  (b) The stream function computed based on the $x$-averaged spanwise DNS velocity field, which was
integrated to obtain the stream function, i.e.,
$\psi_{x_{ave}}(y,z)=-\frac{\partial w_{x_{ave}}^{\prime}}{\partial
y}$. These plots are reproduced from \cite{GMcPBD09}. }
\end{center}
\end{figure*}

This paper is organized as follows.  In the next section, we describe the streamwise constant (so-called 2D/3C) model and the idealized steady-state stream function model representing the cross-stream components of streamwise homogenous features.  This stream function is used as input to a steady-state 2D/3C streamwise momentum equation.  The solution corresponding to each stream-function input can be thought of as a forced solution of the respective streamwise deviation from the laminar flow.  These streamwise velocity fields are compared to a spatial field of DNS data obtained from the Kawamura group \cite{KawData} in order to verify the ability of the model to capture the relevant features of turbulent flow.  We compute the spanwise/wall-normal ($z$--$y$) plane forcing required to produce each of the stream functions described above and study the input-output response from this forcing input to the streamwise velocity's deviation from laminar. The optimal spanwise wavelength computed in this manner is consistent with linear studies.  However, the nonlinearity in the model gives additional insight to the relationship between amplification and the turbulent velocity profile.  In fact, this work demonstrates that there is an important tradeoff between linear amplification mechanisms and the nonlinearity required to develop an appropriately shaped turbulent velocity profile.  The paper concludes with a summary of our results and directions for future work.

\section{Models}

\subsection{The 2D/3C Model} \label{sec:theModel} The 2D/3C model
for plane Couette flow discussed herein is obtained by setting the
streamwise ($x$--direction) velocity derivatives in the full NS equations to zero \citep{BobbaThesis}.  This can be thought of as a projection of the
NS into the streamwise constant space.  The velocity field
is then decomposed into components
$\vec{\mathbf{u}}=[U+u^{\prime}_{sw},V+v^{\prime}_{sw},W+w^{\prime}_{sw}]$;
where $(U, V, W)$ with $U=U(y)=y$, $V=W=0$ is the laminar flow and
$(u^{\prime}_{sw},v^{\prime}_{sw},w^{\prime}_{sw})$ are respectively the
streamwise, wall-normal and spanwise time dependent deviations from laminar in the
streamwise constant sense.  One can explicitly
show that for Couette flow this 2D/3C formulation also results in
a system with zero streamwise pressure gradient.

A stream function $\psi(y,z,t)$, such that
\[ \label{eqn:streamfunction} v^{\prime}_{sw}=\frac{\partial
\psi}{\partial z}; \;\quad w^{\prime}_{sw}=-\frac{\partial
\psi}{\partial y} \] ensures that the resulting model satisfies the
appropriate $2D$ continuity equation.  This yields
\begin{subequations}
\label{eqn:full2D3C}
\begin{align}
 \frac{\partial u^{\prime}_{sw} }{\partial t} &=  - \frac{\partial \psi }{\partial z}\frac{\partial u^{\prime}_{sw} }{\partial y}
 - \frac{\partial \psi }{\partial z}\frac{\partial U}{\partial y}
 + \frac{\partial \psi }{\partial y}\frac{\partial u^{\prime}_{sw}}{\partial z} + \frac{\Delta}{R} u^{\prime}_{sw} \label{eqn:full2D3C-a}\\
 \frac{{\partial \Delta \psi }}{{\partial t}} &=  - \frac{\partial \psi }{\partial z}\frac{\partial \Delta \psi }{\partial y}
 + \frac{\partial \psi }{\partial y}\frac{\partial \Delta \psi }{\partial z} + \frac{1}{R}\Delta ^2 \psi,\label{eqn:full2D3C-b}
  \end{align}
\end{subequations}
where $\Delta=\frac{\partial^2 }{\partial y^2}+\frac{\partial^2
}{\partial z^2}$.  There is no slip or penetration at the
wall and periodic boundary conditions are assumed for the spanwise direction.

The Reynolds number employed for all computations described herein is $R=R_w=\frac{U_w h}{\nu}$, where the
$U_w$ is the velocity of the top plate, $h$ is the channel height
and $\nu$ is the kinematic viscosity of the fluid.  All distances
and velocities are respectively normalized by $h$ and $U_w$.

\begin{figure*}[!t]
\begin{center}
\subfigure[]{\label{fig:SteadyStateContour}
\includegraphics[height=2.25in,width=3.45in,clip]{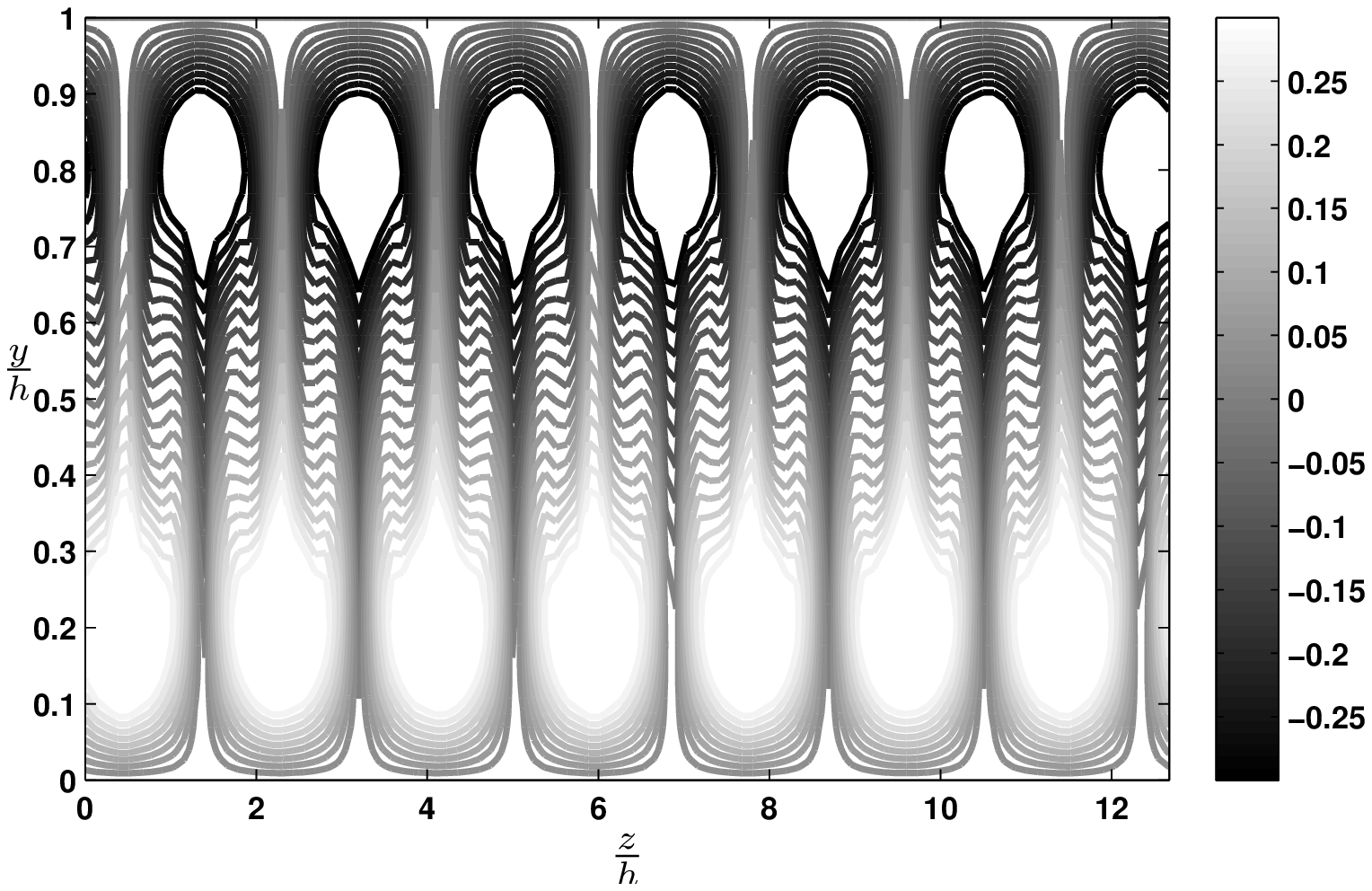}}
\subfigure[]{\label{fig:DNSContours}
\includegraphics[height=2.25in,width=3.45in,clip]{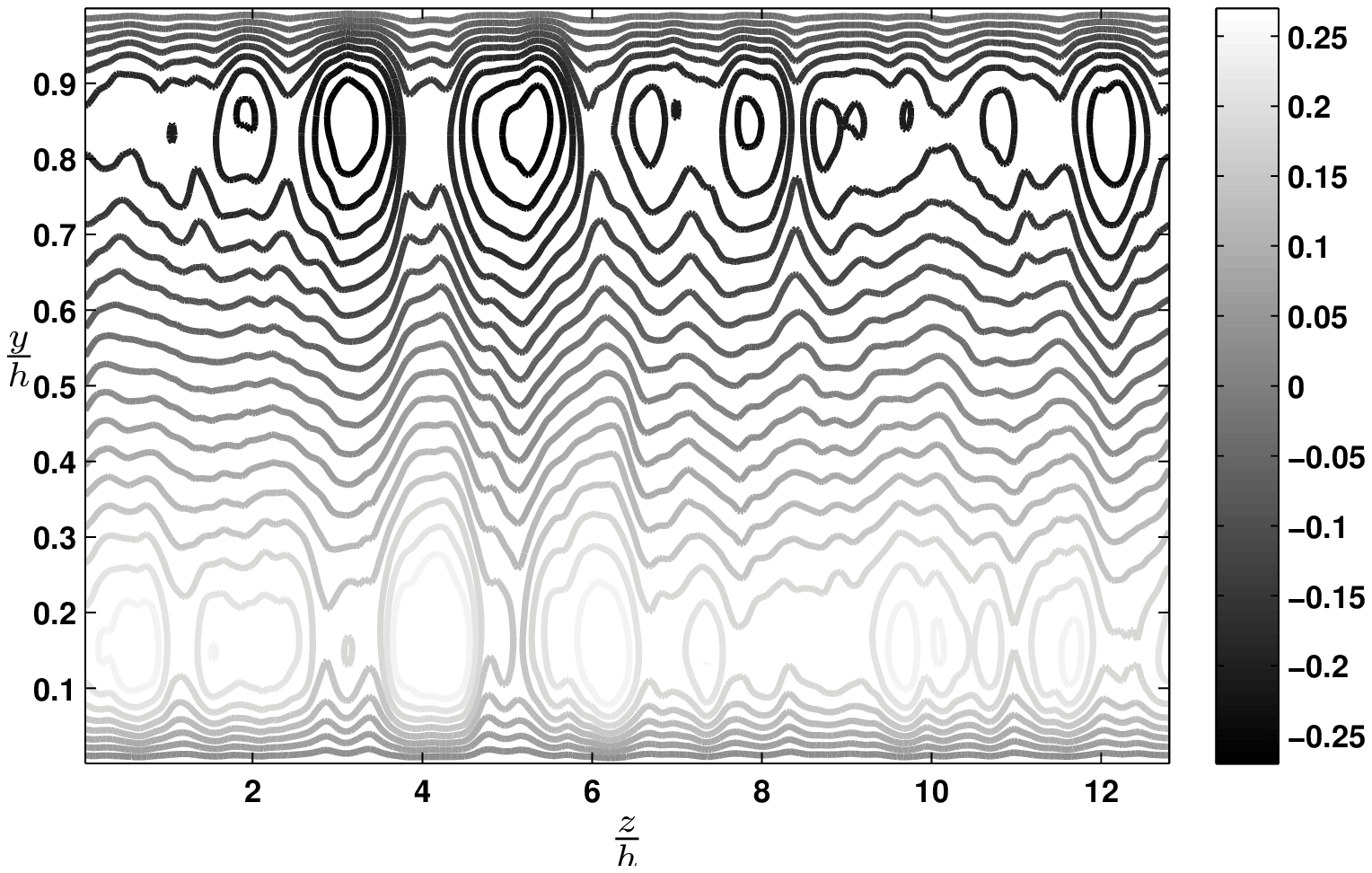}}
\caption{Contour plots of (a)
$u^\prime_{{sw}_{ss}}$, from $\psi_{ss}(y,z)=0.00675\sin ^2 \left(
{\pi y} \right)\cos \left( {\frac{{2\pi }}{{1.82 }}z} \right)$ and
(b) $u^\prime_{x_{ave}}$ the
streamwise velocity component of the $x$-averaged DNS data.  All
plots correspond to $R=3000$ and have the same
contour levels. These plots are reproduced from \cite{GMcPBD09}.}\label{fig:SteadyStatevsDNS}
\end{center}
\end{figure*}

\subsection{The Stream Function Model}

As a first step, we focus on the effect of large-scale streamwise elongated features in the core
of a fully turbulent flow.  We limit our study to cross-stream (i.e. wall-normal/spanwise plane) inputs because energy amplification of perturbations (forcing) from the wall-normal and spanwise directions have been shown to scale as $R^3$ whereas all of the other input-output combinations admit only $R$ scaling \cite{JB05}. We focus on the effect of these cross-stream inputs on the streamwise component of the flow.

We are interested in developing a simple analytic model for the steady-state stream functions $\psi_{ss}(y,z)$ that will define our inputs.  This will lead to computational tractability and better lends itself to analytical studies.  In Barkley and Tuckerman \cite{BT07} it was shown that laminar-turbulent flow patterns in plane Couette flow can be reproduced using a cross-stream stream function of the form $\psi(y,z)=\psi_0(y)+\psi_1(y)\cos(k_z z)+\psi_2(y) \sin(k_z z)$.  We use this study as guidance but set the zeroth-order term to
zero because a nonzero $\psi_0$ produces a nonzero-mean spanwise flow $w^\prime_{ss}$, which is not representative of the velocity field we are interested in studying.  This leads to the following simple doubly harmonic function
\begin{equation}
\label{eqn:psimodel}
 \mathbf{\psi_{ss}} (y,z) = \varepsilon\sin ^2 \left( {q\pi y} \right)\cos
\left(\frac{2\pi
}{\lambda_z }z \right),
\end{equation}
which obeys the
boundary conditions \cite{GMcPBD09,GaymeThesis}.

In order to estimate the values for $\lambda_z$, $\varepsilon$ and $q$ we averaged a full spatial field of DNS data \cite{KawData} along the $x$--component to obtain an approximation of a streamwise constant flow field. In the sequel, we refer to the this streamwise averaged DNS field as the $x$--averaged DNS data. A full discussion of this velocity field and its use as an approximation for streamwise constant data is given in \citep{GaymeThesis,GMcPBD09}.  The $x$--averaged flow fields were used to ensure that our $\psi_{{ss}}$ model as well as the corresponding wall-normal and spanwise velocities, have the correct features.

Figure \ref{fig:Psi_from_w0} shows $\psi=-\frac{\partial w_{x_{ave}}^{\prime}}{\partial y}$ computed by integrating the $x$--averaged spanwise velocity ($w{^\prime}_{x_{ave}}(y,z)$) from the DNS data at $R=3000$.  The $\psi_{ss}$ model corresponding to equation (\ref{eqn:psimodel}) with $q=1$, $\varepsilon=0.00675$ selected to match the magnitude of the integrated $w_{x_{ave}}^{\prime}$ and $v_{x_{ave}}^\prime$ DNS fields and $\lambda_z\approx 1.8$ selected to match the DNS' fundamental spanwise wavelength is provided in Figure \ref{fig:Psi-a}.  Comparison of figures  \ref{fig:Psi_from_w0} and \ref{fig:Psi-a} indicate that the $\psi_{ss}$ model shows good agreement with the DNS data approximation in the region of highest signal.

\section{Forced Solutions}
\label{sec:forced_solns}

The steady-state solutions of \eqref{eqn:full2D3C-a} corresponding to steady-state stream functions of the form \eqref{eqn:psimodel} are of interest for two reasons. First, they allow us to isolate the nonlinear streamwise velocity equation in order to demonstrate that its nonlinear coupling filters an appropriately constructed $\psi_{ss}(y,z)$ towards the expected ``S'' shape of the turbulent velocity profile.  It also gives insight into mechanisms that create the momentum (energy) transfer which generates this blunted profile.

The forced solutions shown herein are presented solely to demonstrate that the 2D/3C model (\ref{eqn:full2D3C}) is representative of certain aspects of turbulent behavior and as such, an amplification study based on this model is of interest.  Gayme et al. \cite{GMcPBD09} and Gayme \cite{GaymeThesis} provide a detailed exploration of the extent to which the 2D/3C equations \eqref{eqn:full2D3C} can be used as a model for turbulent behavior.  For all of the results presented in this section we solved for the forced solution $u^{\prime}_{{sw}_{ss}}(y,z)$ using both a least-squares approach and iteratively using an explicit Euler method for comparison.  The initial studies were carried out using the same grid resolution as in the DNS data described in \cite{KawData}, (i.e. using a $96\times 512$ grid on the $y$--$z$ plane).  We then reduced the resolution to a $y$--$z$ plane grid of $48\times 100$.  We found negligible differences in the results between these two grid sizes. In the sequel, we only report the results for the $48\times 100$ grid and the explicit Euler iterative solution.

\subsection{The Velocity Field}

Figure \ref{fig:SteadyStateContour} shows a contour plot of $u^{\prime}_{{sw}_{ss}}(y,z)$ resulting from a stream function $\psi_{ss}$
with the same parameters as in Figure \ref{fig:Psi-a},  (i.e. $q=1$, $\varepsilon=0.00675$ and $\lambda_z=1.8$, at $R=3000$).  This figure shows that our computed streamwise deviation from the laminar velocity has features consistent with the difference between a laminar and turbulent velocity field.  For example, the velocity gradients near the walls that are associated with an ``S'' shaped (or blunted) turbulent velocity profile.  Figure \ref{fig:SteadyStateContour} also shows good qualitative agreement with the $x$--averaged DNS data (shown in Figure \ref{fig:DNSContours} with the same contour levels as \ref{fig:SteadyStateContour}).  This DNS data has previously been shown to have a spanwise mean velocity corresponding to the full turbulent velocity profile \cite{GMcPBD09}.  The steady-state streamwise velocity deviation from laminar $u^{\prime}_{{sw}_{ss}}(y,z)$ has similar structural features to the DNS data in that both have near-wall minimum and maximum peaks that are out of spanwise phase with one another top-to-bottom.  There is however, more variation in the deviation from laminar across the span as compared to the DNS field.

\subsection{Mean Profile}
\label{sec:vel_profiles}

\begin{figure}
\begin{center}
\includegraphics[height=2in,width=3.25in,clip]{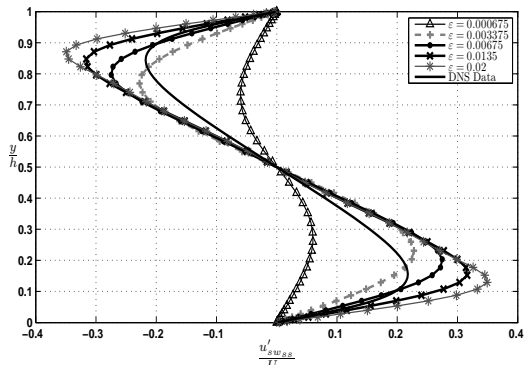}
\caption{Variation of the
2D/3C (streamwise constant) deviation from laminar, $u'_{sw_{ss}}$
with perturbation amplitude ($\varepsilon$); based on input $\psi_{ss}(y,z)=\varepsilon \sin ^2 \left( {\pi y} \right)\cos
\left( {\frac{{2\pi }}{{1.82 }}z} \right)$.  This figure is reproduced from \cite{GMcPBD09}. }
\label{fig:Amp_var}
\end{center}
\end{figure}

The effective energy redistribution through the forced streamwise velocity deviations is further investigated through a spanwise mean over the streamwise velocity deviation from laminar.  This is carried out at five perturbation amplitudes ($0.000675 \leq\varepsilon\leq 0.02$), all at $R=3000$.  Figure \ref{fig:Amp_var} shows averages across the span of $u^{\prime}_{sw_{ss}}(y,z)$ for these five $\varepsilon$ values along with a similar average of the $x$--averaged streamwise velocity field of the DNS.  The spanwise average of the DNS, has been validated against other results in the literature by Tsukahara et al. \citep{KawData}.  The use of $\psi_{ss}$ from (\ref{eqn:psimodel}) as an input to a the steady-state \eqref{eqn:full2D3C-a} produces streamwise velocity profiles whose shapes are consistent with the $x$--averaged DNS data.  The
peaks are, however located at different wall-normal positions.

The fact that an exact match (with DNS data) for the wall-normal peak position is not obtained is not unexpected given the simplicity of the wall-normal variation in the steady-state model, as well as the streamwise constant and steady-state assumptions. Clearly the full turbulent field is neither
streamwise constant nor steady-state.  The main point of presenting the velocity field arising from the forced steady-state model is to illustrate its effectiveness in reproducing the momentum redistribution associated with the change in the velocity profile from laminar to turbulent. This means that \eqref{eqn:full2D3C} provides more information about the turbulent velocity field than a linear model, (which cannot produce the change in mean velocity profile between laminar and turbulent flows).  Therefore studying input-output amplification in this model may provide us with some additional insight compared to the traditional analysis performed using linear models.

The simple steady-state model described herein reasonably predicts the essence of the mean behavior at the expense
of losing some of the smaller scale details. For example, an exact characterization of the wall-normal
variation and, of course, the small-scale turbulent velocity fluctuations are not captured in this analysis. These
results suggest that the phenomenon that is responsible for blunting of the velocity profile in the mean sense is a direct consequence of
the interaction between rolling motions caused by the $y$--$z$ stream function and the laminar profile.  In other words, this study
provides strong evidence that the nonlinearity needed to generate the turbulent velocity profile is dominated by the nonlinear terms that
are present in the $u^{\prime}_{sw}(y,z,t)$ evolution equation (\ref{eqn:full2D3C-a}).

\section{Input-Output Amplification}
\label{sec:energy}

\subsection{Energy Amplification}
\label{sec:energyamp}

In order to discuss input-output amplification, it is useful to determine the forcing
required to produce a steady-state $\psi_{ss}$.  This is accomplished by solving a forced version of the steady-state $\psi$
evolution equation in (\ref{eqn:full2D3C-b}) for some forcing term $\eta_{ss}(y,z)$.  The linearized version of this forcing, which by abuse of notation we also refer to as $\eta_{ss}(y,z)$, is given by
\begin{equation}
\label{eqn:lin_eta}
 \eta_{ss} (y,z) =- \frac{1}{R }\Delta ^2 \psi_{ss}.
\end{equation}
This $\eta_{ss}$ can be viewed as the deterministic forcing required to produce a
particular $\psi_{ss}$.  In
the sequel, we use the linear $\eta_{ss}$ of Equation
(\ref{eqn:lin_eta}) for all of the computations.  For a complete
discussion of the use of a linear $\psi$ equation see
\cite{GMcPBD09, GaymeThesis}.

The input-output response can now be studied through an amplification factor of the form
\begin{equation}
\label{eqn:Amp_factor} \Gamma_{ss}  = \frac{{\left\| {u^\prime_{sw_{ss}} }
\right\|^2 }}{{\left\| \eta_{ss}  \right\|^2 }}.
\end{equation}
$\Gamma_{ss}$ is a nonlinear analog of the type  $L_2$--to--$L_2$ induced
norm that has been used to study the optimal response of the LNS with harmonic input/forcing, see for example \cite{SH_bk}.   The energy in (\ref{eqn:Amp_factor}) is
defined in terms of the squared $2$-norm.  For each
$2$-dimensional component $\beta(y,z)$ this quantity is
\begin{equation}
\label{eqn:norm}
\begin{aligned}
\|\beta\|^2=& \int_{z_1}^{z_2} \int_{0}^{1} \beta(y,z)^2\,dy \;dz
\\ \approx& \frac{\Delta z\Delta y}{ L_y\;L_z}
\sum_{k=1}^{N_z-1}\sum_{j=1}^{N_y-1}\beta^2(y_{j},z_{k}).
\end{aligned}
\end{equation}
where $\Delta y=y_2-y_1$ and  $\Delta z=z_2-z_1$ respectively
represent the space between the $z$ and $y$ grid points.

\begin{figure*}[!t]
\begin{center}
\subfigure[]{\label{fig:VaryR-a}
\includegraphics[width=3in, height=2in ,clip]{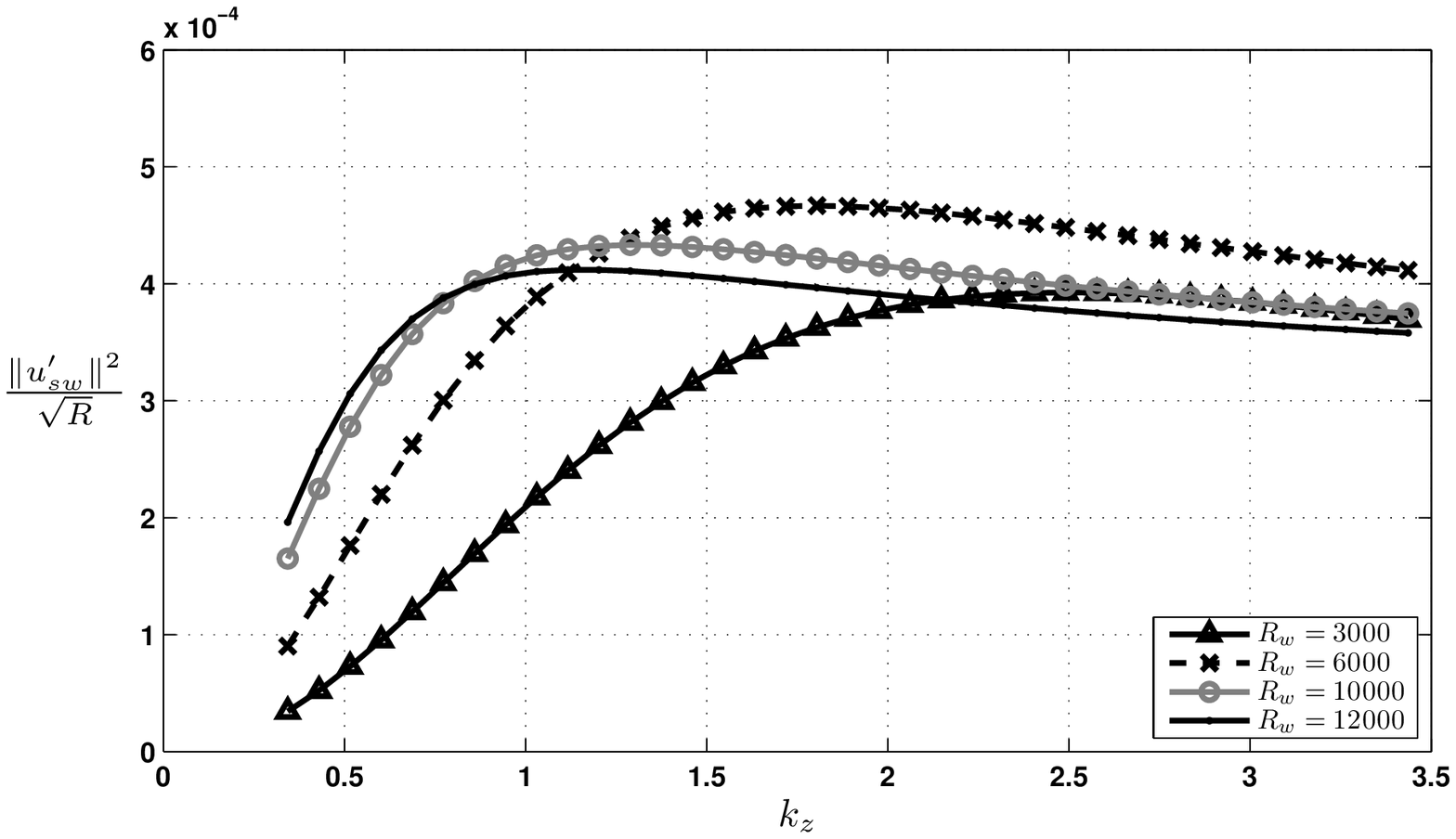}}
\subfigure[]{\label{fig:VaryR-b}
\includegraphics[width=3in, height=2in,clip]{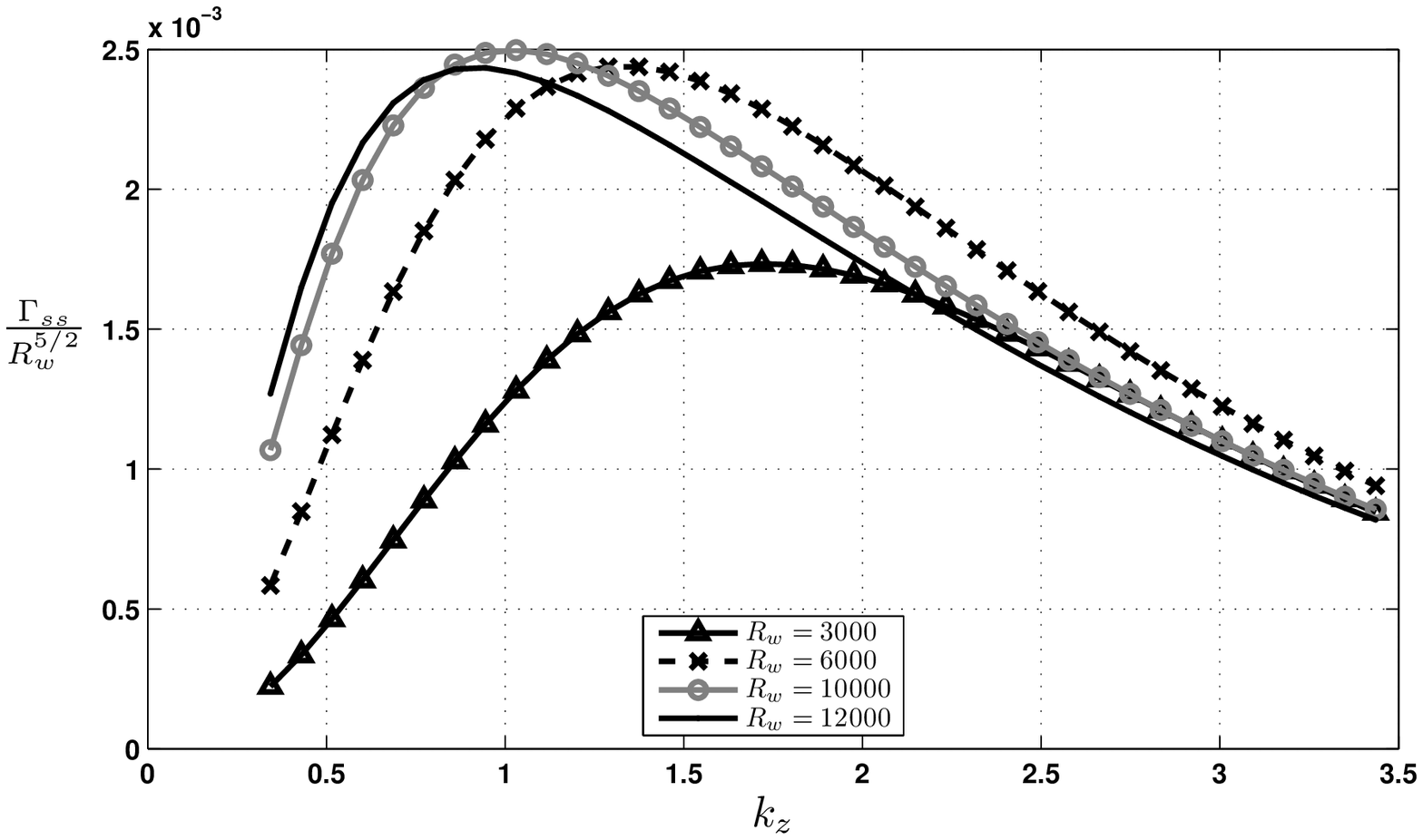}}
\subfigure[]{\label{fig:VaryAmp-a}
\includegraphics[width=3in, height=2in,,clip]{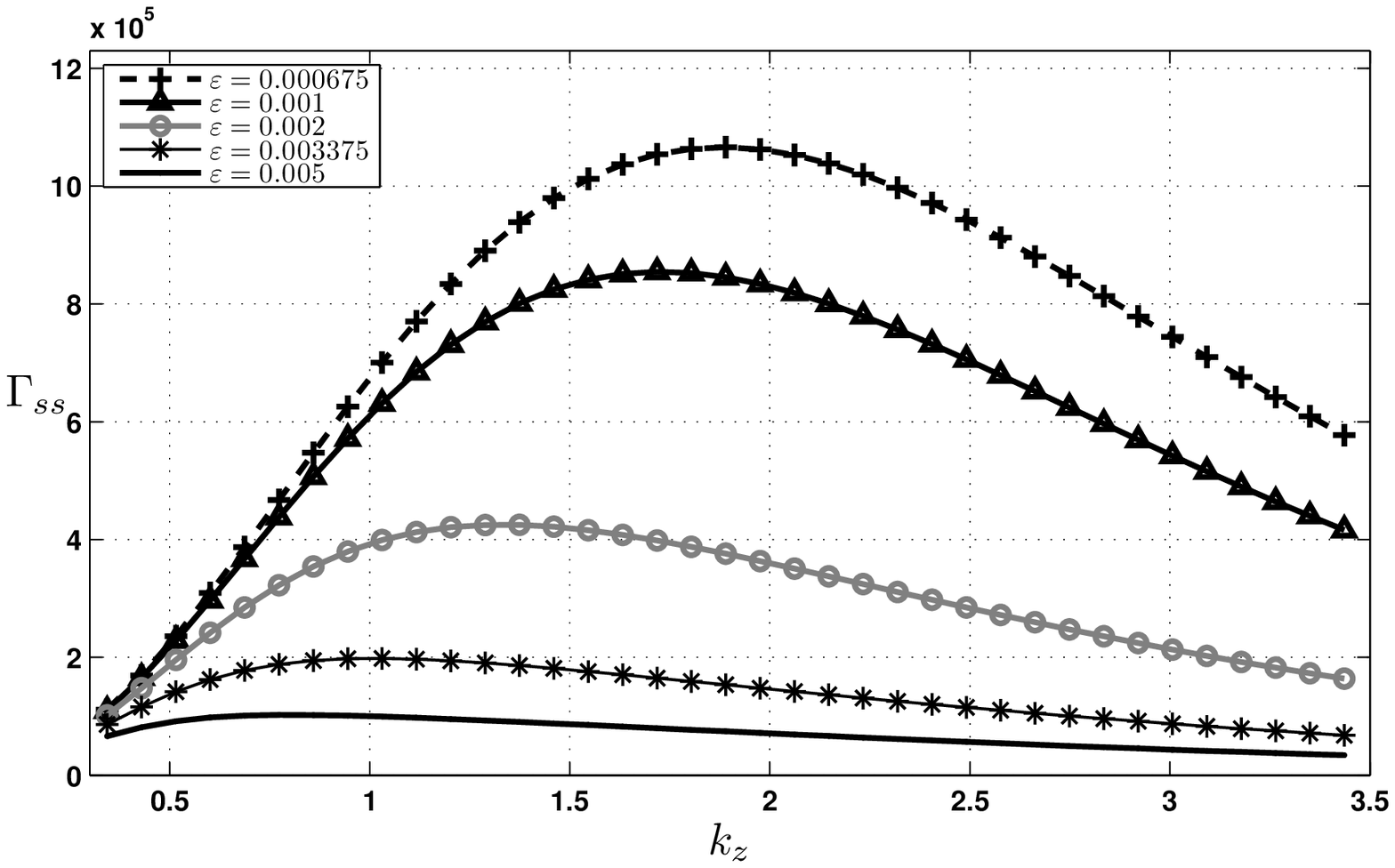}}
\subfigure[]{\label{fig:VaryAmp-b}
\includegraphics[width=3in, height=2in,,clip]{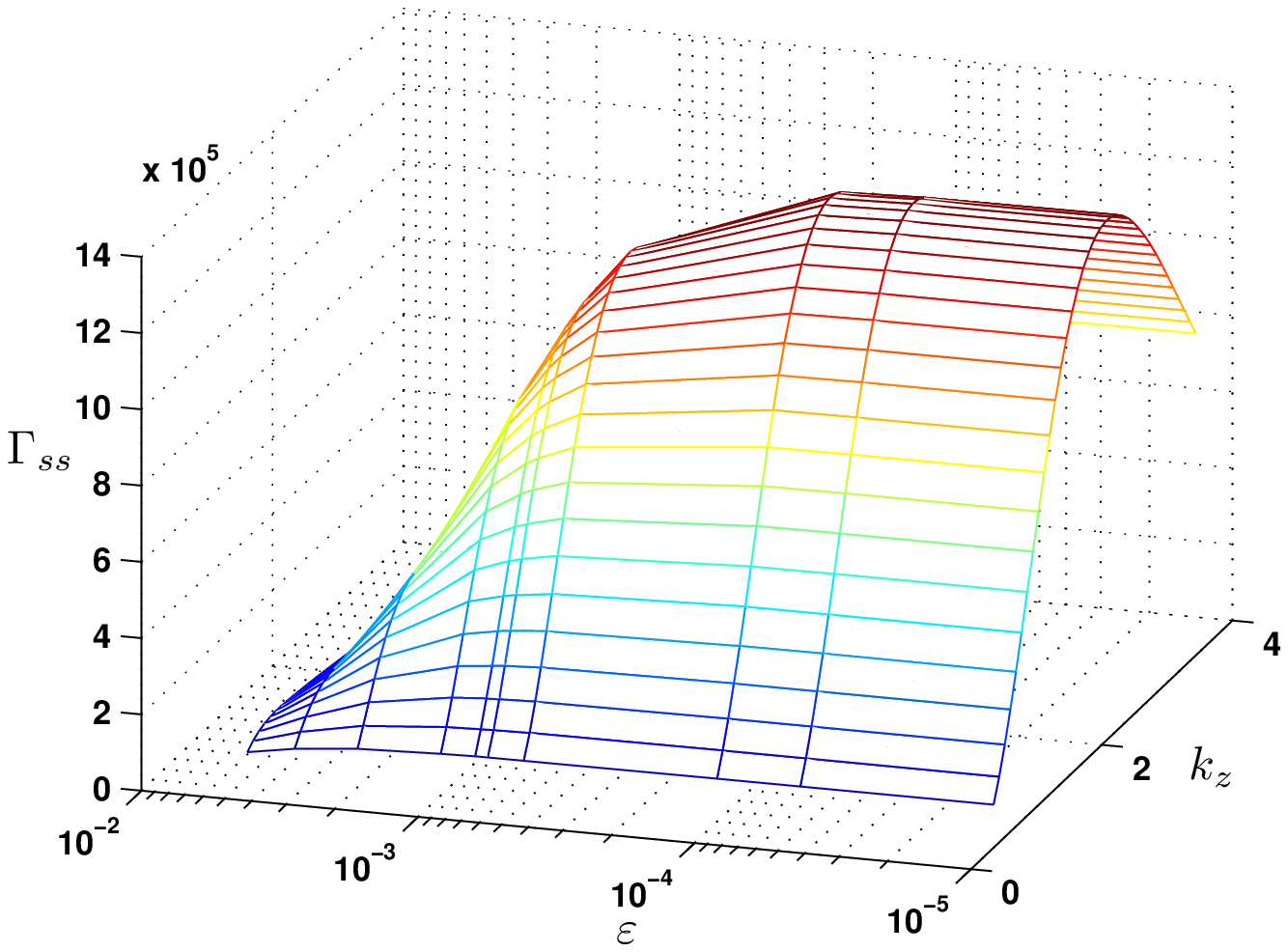}}
\caption{(a) The streamwise energy scales as $\sqrt{R}$.
(b) The amplification factor $\Gamma_{ss}$ scales as
$R^{\frac{5}{2}}$. The optimal spanwise wavenumber occurs at the
maximum $\Gamma_{ss}$ for each $R$. (c, d) $\Gamma_{ss}$ for
different values of $\varepsilon$ all at $R=3000$.  Both
$\Gamma_{ss} $ and the optimal spanwise wavenumber monotonically
decrease with increasing $\varepsilon$. }\label{fig:VaryR}
\end{center}
\end{figure*}

\subsection{Reynolds Number Scaling}

The scaling of $u_{sw_{ss}}^{\prime}$ with $R$ for a particular $\varepsilon$ is unclear from equations \eqref{eqn:full2D3C-a} and \eqref{eqn:psimodel}.  An empirical relationship
was computed using the stream function model
(\ref{eqn:psimodel}) with $q=1$ and $\varepsilon=0.001$ for four
different values of $R$: $3000$, $6000$, $10000$ and $12000$.
Figure \ref{fig:VaryR-a} shows that $\|u_{sw_{ss}}^{\prime}\|^2$ appears to scale as a function of ${\sqrt{R}}$ at the higher wavenumbers ($k_z >2$), for the $R$ values selected.  The energy peaks also seem to collapse under this scaling for the higher Reynolds numbers we considered.

The scaling of $\Gamma_{ss}$ can be estimated by combining this ${\sqrt{R}}$ scaling of
$\|u_{sw_{ss}}^\prime\|^2$ with the $\frac{1}{R}$ scaling of
$\eta_{ss}(y,z)$, as seen in \eqref{eqn:lin_eta}.  Thus, $\Gamma_{ss}$ should scale as a function
of $R^{\frac{5}{2}}$.  Figure \ref{fig:VaryR-b}, which shows $\frac{\Gamma_{ss}}{R^{5/2}}$ for the same
$R$'s and $\varepsilon$ indicates that the amplification factor data does collapse well under the $R^{5/2}$ scaling, especially at
the higher wave numbers.  However, the $R=3000$ data peak does not seem to follow this relationship.  This discrepancy can be explained by looking at previous linear studies.

The scaling of the input-output amplification for streamwise constant disturbances of the LNS equations was expressed as
$f_1(k_z)R+f_2(k_z)R^3$ by Bamieh and Dahleh~\cite{BD01}.  Furthermore, they reported that the form of $f_1(k_z)$ and $f_2(k_z)$ means that the linear term dominate at low Reynolds numbers~\cite{BD01}.  The difference in scaling over different Reynolds number ranges was confirmed by a low Reynolds number linear study of
Poiseuille flow that showed energy amplification at $k_x=0$ scales with
$R^{\frac{3}{2}}$ for the range $800\leq R\leq 5000$ and $R^3$ for
larger Reynolds numbers \cite{FI94}. In that study, $R$ was
normalized on half channel height $\delta$.  The equivalent
normalization would make our Reynolds number range $750\leq R_c\leq
3000$.  Optimal amplification
studies based on initial conditions also support $R$ scaling at
low Reynolds numbers \cite{FI94}.  Based on these studies, both the fact that our
scaling is less than $R^3$ and that we found a lower $\Gamma_{ss}$ peak value for $R=3000$ (corresponding to $R_c=750$) are not unreasonable given the low
Reynolds numbers we are employing.

\subsection{Optimal Spanwise Spacing}
\label{sec:Opt_Space}

Figure \ref{fig:VaryR-a} indicates that $\|u_{sw_{ss}}^{\prime}\|^2$
increases with $k_z$ until it reaches a maximum value and then
levels off.  We can similarly find a relationship between $k_z$ and
$\Gamma_{ss}$ by substituting the expression for
$\mathbf{\psi_{ss}}$ from (\ref{eqn:psimodel}) into the linearized
noise equation (\ref{eqn:lin_eta})
\begin{equation}
\label{eqn:noise}
\begin{aligned}
 \eta_{ss}(y,z)=
\frac{1}{R}\cos(k_z
z)\biggl\{\left(8 q^4 \pi ^4+2 q^2 k_z^2 \pi ^2\right)+\\
\left[4q^2k_z^2\pi^2-\left(k_z^2+4q^2\pi^2\right)^2
\right]sin^2(q\pi y)\biggr\}.
\end{aligned}
\end{equation}
Equation \eqref{eqn:noise} illustrates that the noise scales with $k_z^4$ and $q^4$.
So, the forcing
energy $\|\eta_{ss}\|^2$ monotonically increases with $k_z$
while $\|{u^\prime_{sw_{ss}} }\|^2$ peaks and then levels off. Thus, even though larger $k_z$ is associated with higher
forcing the corresponding amplification factor does not continue to
increase. There is an optimal $k_z$ that generates the most
amplification: This is the dominant wavenumber corresponding to
optimal spanwise spacing.  In this section we explore how changes in $\varepsilon$ and $R$ relate to the optimal spacing and discuss how our results compare to what has been previously reported in the literature.

The peak values of $\Gamma_{ss}$ for the Reynolds numbers considered
in Figure \ref{fig:VaryR-b} correspond to spanwise wave numbers of
$k_z=0.86$, $1.0$, $1.4$ and $1.7$; for $R=12000,\,10000,\, 6000$
and $3000$ respectively. This amounts to wavelengths of $7.3h$,
$6.1h$, $4.6h$ and $3.7h$.  To determine if these are optimal values over the entire parameter set we need to determine the relationship between $\Gamma_{ss}$ and
amplitude $\varepsilon$.

Figure \ref{fig:VaryAmp-a} shows $\Gamma_{ss}$ for an amplitude
range of $0.000675\leq\varepsilon\leq 0.005$ all with $R=3000$.  Over most of the range both $\Gamma_{ss}$
and the optimal spanwise wavenumber monotonically decrease with
increasing $\varepsilon$, although there appears to be a collapse at the minimal
wavelengths.  Therefore, the peak $\Gamma_{ss}$
over all the amplitudes we selected (i.e. the optimal $k_z$) occurs at the lowest $\varepsilon$.
Figure \ref{fig:VaryAmp-b} shows that continuing to reduce $\varepsilon$ results in convergence to an
optimal wavenumber of $k_z=2.06$, which corresponds to $\lambda_z=3.05h$, for all $\varepsilon\leq 0.0001$. We obtain the same optimal wavelength when we repeat this procedure for
$R= 6000,\,10000$ and $12000$.

Much of the literature, (e.g., \cite{FI93b,BF92,G91})
related to optimal spanwise spacing has shown optimal spanwise wave numbers $k_z\in~[2.8,4]$.  However, in many cases these studies were aimed at
determining spanwise spacing in the near wall (inner scaled) region.  Recent Poiseuille flow studies using the LNS linearized about a
turbulent velocity profile, where an eddy viscosity is used to
maintain the profile, found that at high Reynolds numbers there are actually
two peaks in the optimal energy growth curves, one scaling in inner
units and the other in outer units \cite{dAJ06,PGCD09}.  The outer
unit peak appears to correspond to the large-scale structures
that have a spanwise spacing of approximately
$\lambda_z\in[2, 5.2]\delta$.

The only Couette flow study to look at both inner and outer unit
scalings reported results at $R=3000$ \cite{HC10}.  At this low Reynolds number there is little to no scale separation between the peaks.
They studied several input-output response functions and found that the optimal
response to harmonic forcing occurs when $\lambda_z= 3.85h$.  This harmonic forcing study is more closely related to our analysis than the initial condition-based studies reported in most of the other work.
Couette flow DNS \cite{KLJ96} and experimental studies \cite{KU08} have observed large-scale (outer region) feature spacing in the range $k_z=[2,2.55]h$. Thus, our results ($\lambda_z=3.05h$) lie right in between those of the analytical Couette flow study \cite{HC10} and the flow field observations. A constant optimal wavelength across Reynolds numbers is also consistent with previous studies using a linear model with an eddy viscosity based turbulent velocity profile \cite{dAJ06,PGCD09}.

\subsection{Mean Velocity Profile versus Optimal $k_z$}

\begin{figure*}
\begin{center}
\subfigure[]{\label{fig:VaryAmp_profiles}
\includegraphics[width=0.475\textwidth,clip]{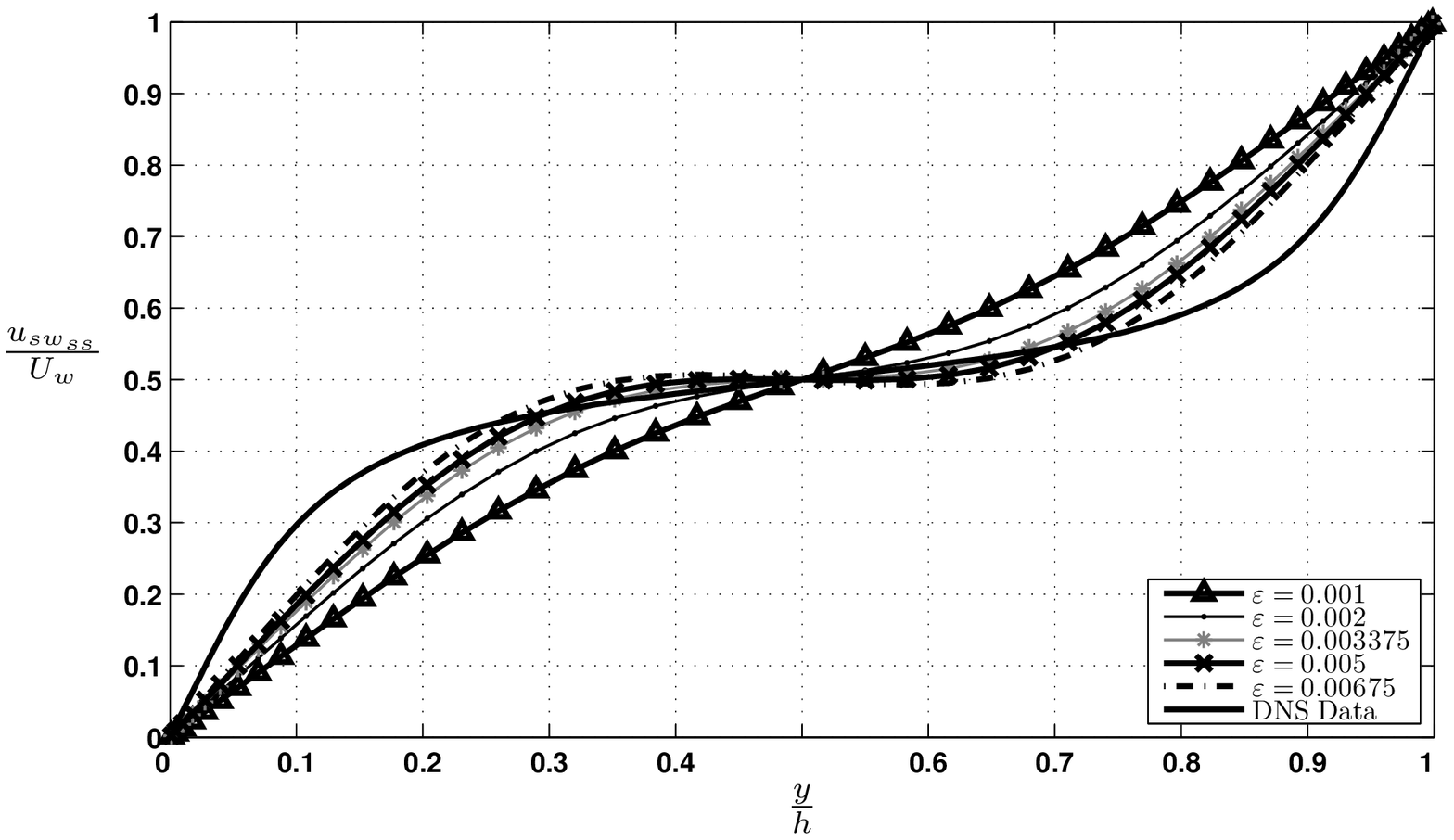}}
\subfigure[]{\label{fig:mean_vary_kz}
\includegraphics[width=0.475\textwidth,clip]{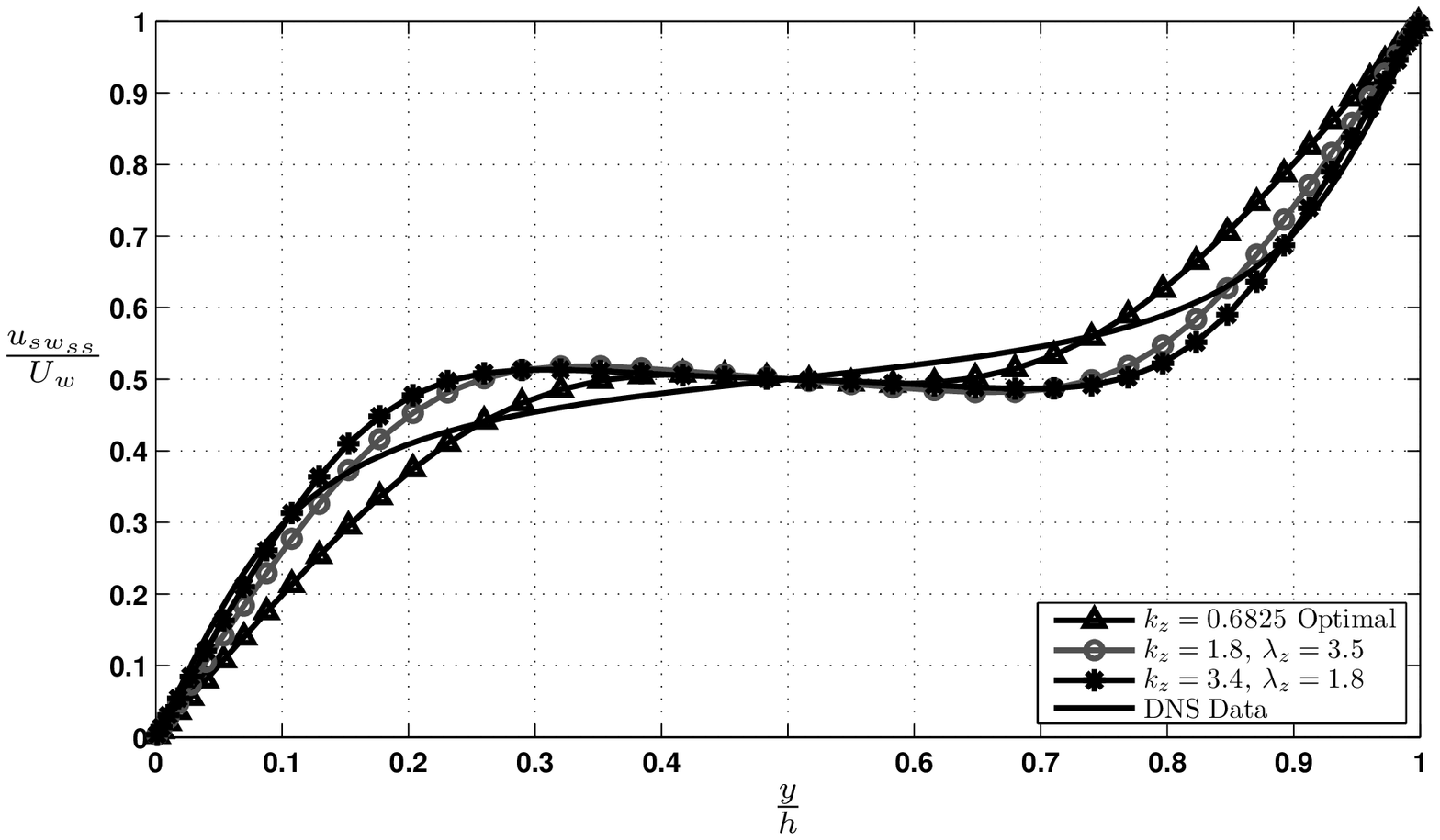}}
\caption{(a)The mean velocity profile of DNS data along with one computed from the steady-state \eqref{eqn:full2D3C-a} for a $\psi_{ss}$ model \eqref{eqn:psimodel} with $q=1$, over a range of $\varepsilon$ with $k_z$ corresponding to the peak $\Gamma_{ss}$ for each $\varepsilon$ considered. (b) The mean velocity profile for $\varepsilon=0.00675$ at a number of different $k_z$ values compared with DNS data.  The data in both (a) and (b) correspond to $R=3000$. }
\end{center}
\end{figure*}

\begin{figure}

\includegraphics[width=0.475\textwidth,clip]{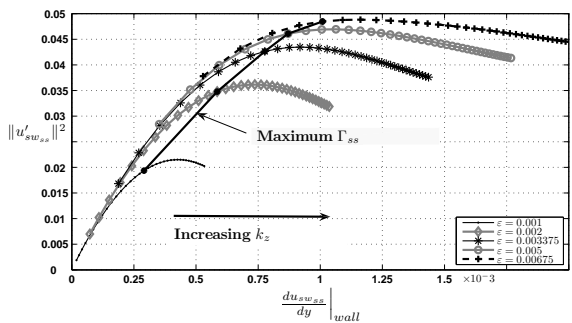}
\caption{The velocity gradient at the wall continues to increase while both
$\Gamma_{ss}$ and the energy $\|u_{sw_{ss}}^\prime\|^2$ peak and
then drop off. The solid black line represents the peak
$\Gamma_{ss}$ for each $\varepsilon$.}\label{fig:shear}
\end{figure}

Figure \ref{fig:VaryAmp_profiles} shows the steady-state mean velocity profile computed using $\psi_{ss}$ models \eqref{eqn:psimodel} with amplitudes
in the range $0.001\leq\varepsilon\leq 0.00675$ at their corresponding optimal values of $k_z$ along with the
DNS data.  All plots correspond to $R=3000$.  The velocity gradients at the wall increase with $\varepsilon$.  The fit in the center of the channel also approaches the DNS data as $\varepsilon$ increases, although at $\varepsilon=0.005$ and above the curve overshoots the DNS.  There is no amplitude that exactly matches the DNS data and the fit is especially bad in the near-wall region.  As previously discussed, this is because the assumptions inherent in the 2D/3C model neglect the smaller scale activity that dominates in the near-wall region.  In the core, the mean velocity curves, $\frac{u_{sw_{ss}}}{U_w}$, corresponding to $\varepsilon=0.005$ and $\varepsilon=0.00675$ respectively cross the DNS curve at a $y^+\approx 30$ and $y^+\approx 27$, based on the DNS
viscous units. The maximum overshoot in the core (defined by $y^+>30$ in DNS viscous units) is $3.6\%$ and $6.2\%$, respectively
for $\varepsilon=0.005$ and $\varepsilon=0.00675$.  This is remarkably good for such a simplified steady-state model.

The fact that the $\varepsilon=0.005$ and $\varepsilon=0.00675$ mean velocity profiles show the best agreement (most blunting) with the DNS data is not consistent with the fact that maximum amplification occurs at the smallest amplitudes (i.e. $\varepsilon\leq0.0001$).  In order to study this further we looked at different $k_z$ values corresponding to a $\psi_{ss}$ model amplitude of $\varepsilon=0.005$.  Figure \ref{fig:mean_vary_kz} shows the mean velocity profile of the
DNS along with mean velocities for $\varepsilon=0.005$ at the maximum $\Gamma_{ss}$ (optimal wavenumber $k_z=\frac{2\pi}{\lambda_z}=0.69$), at $k_z=1.8$ and at $k_z=3.4$. This last value coincides with $\lambda_z=1.8$, i.e., the
value corresponding to the dominant wavenumber of the $x$--averaged DNS data \cite{KawData} and the results discussed in Section \ref{sec:forced_solns}.  Again, while by definition the amplitude of $\Gamma_{ss}$ is larger for the optimal wavenumber $k_z=0.6825$, the velocity profile has larger velocity gradients at the wall for higher values of $k_z$.  This continued increase in shear stress at the wall as both $k_z$ and $\varepsilon$ increase is better seen in Figure \ref{fig:shear} which depicts the energy in $u^\prime_{ss}$ versus the shear stress at the wall.

Small $\psi_{ss}$ model \eqref{eqn:psimodel} amplitudes, $\varepsilon$'s, correspond to lesser nonlinear coupling between the equations.  On the other hand, at higher $\varepsilon$ the nonlinear terms have a larger magnitude because $\varepsilon$ directly multiplies each of the nonlinear terms.  As $\varepsilon$ decreases the energy amplification increases but the velocity profile blunting decreases (i.e., the profiles become increasingly laminar-like). This behavior can be interpreted as follows; the amplification is dominated by linear mechanisms, whereas the blunting comes from nonlinear interactions.  Moreover, there is some tradeoff between energy amplification and the creation of a turbulent-like blunted mean velocity profile.  The observation that blunting continues to increase with wavenumbers beyond the energy optimal wavenumber as depicted in Figure \ref{fig:shear} appears to indicate that the exact relationship depends on the spanwise wavenumber.  This type of dependence is consistent with linear amplification theory. Further understanding of this tradeoff may provide important insight into the mechanisms associated with both transition and fully turbulent flow.

\section{Conclusions and Future Work}

A simple cross-stream model of large-scale streamwise elongated structures nonlinearly coupled through a steady-state 2D/3C streamwise momentum equation allows us to isolate important mechanisms involved in determining the shape of the turbulent velocity profile.  The momentum redistribution that produces features consistent with the mean characteristics of fully developed turbulence appear to be directly related to the 2D/3C nonlinear coupling in the streamwise velocity evolution equation.  The steady-state 2D/3C model produces a blunted, turbulent-like profile using very simple stream functions. This behavior appears to be robust to small changes in the stream function model.  This repeatability suggests a preference for redistribution of momentum along the wall-normal direction.  Further understanding of the underlying dynamics of this mechanism may provide insight into the transition problem and allow better design of turbulent suppression flow control algorithms.

An input-output analysis in this framework not only provides results consistent with previous studies but also illuminates an interesting interaction between energy amplification and the increased velocity gradient at the wall associated with the turbulent profile.  Essentially, although the input-output amplification monotonically decreases with increasing forcing amplitude, the velocity profiles become increasingly more blunted.  Thus, there is likely a tradeoff between the linear amplification mechanisms and nonlinear blunting mechanisms that determine the nature of the turbulence-like phenomena modeled by \eqref{eqn:full2D3C}.  This tradeoff may have important implications for flow control techniques that target skin friction or the mean profile.

A natural extension of this work would be to refine our stream-function model through adding additional wall-normal and spanwise harmonics to the existing form of Equation \eqref{eqn:psimodel}.  Developing an entirely new model using models of real sources of flow disturbances, as discussed in \cite{GMcPBD09}, may also provide guidance in determining a better $\psi_{ss}$ as a forced steady-state solution of \eqref{eqn:full2D3C-b}.

\section{Acknowledgements}

The authors would like to thank H. Kawamura and T. Tsukahara for providing us with their DNS data.  This research is partially supported by AFOSR (FA9550-08-1-0043) and B.J.M. gratefully acknowledges support from NSF-CAREER award number 0747672 (program
managers W. W. Schultz and H. H. Winter).

\bibliography{../turbulence_roughamp}
\end{document}